\documentclass[iop]{emulateapj-rtx4}
\usepackage{amsmath}
\usepackage{graphicx}
\usepackage{natbib}
\usepackage{placeins}

\def\bfnabla{{\mbox{\boldmath $\nabla$}}}

\renewcommand\bv{{\mbox{\boldmath $v$}}}

\newcommand\bP{{\mbox{\boldmath $P$}}}
\newcommand\bF{{\mbox{\boldmath $F$}}}
\newcommand\bfr{{\sf\boldmath f}}
\newcommand\bI{{{\sf\boldmath I}}}

\newcommand\bg{{\mbox{\boldmath $g$}}}

\newcommand\Crat{\mathbb{C}}
\newcommand\Prat{\mathbb{P}}

\newcommand{\odif}[2]{\ensuremath{ \frac{d #1}{d #2}}}
\def\<{\,\langle\langle}
\def\>{\,\rangle\rangle}

\begin{document}

\title{Non-linear Evolution of Rayleigh-Taylor Instability in a Radiation Supported Atmosphere}

\author{Yan-Fei Jiang\altaffilmark{1}, Shane W. Davis\altaffilmark{2} and James M. Stone\altaffilmark{1}}
\altaffiltext{1}{Department of Astrophysical Sciences, Princeton
University, Princeton, NJ 08544, USA} 
\altaffiltext{2}{Canadian Institute for Theoretical Astrophysics. Toronto, ON M5S3H4, Canada} 

\begin{abstract}

The non-linear regime of Rayleigh-Taylor instability (RTI) in a
radiation supported atmosphere, consisting of two uniform fluids with 
different densities, is studied numerically.  We perform
simulations using our recently developed numerical algorithm for
multi-dimensional radiation hydrodynamics based on a variable
Eddington tensor as implemented in Athena, focusing on the regime
where scattering opacity greatly exceeds absorption opacity.  We find
that the radiation field can reduce the growth and mixing rate of RTI,
but this reduction is only significant when radiation pressure
significantly exceeds gas pressure. Small scale structures are also
suppressed in this case.  In the non-linear regime, dense fingers sink
faster than rarefied bubbles can rise, leading to asymmetric
structures about the interface.  By comparing the 
calculations that use a variable Eddington tensor (VET) 
versus the Eddington approximation, we demonstrate that
anisotropy in the radiation field can affect the non-linear development of
RTI significantly.  We also examine the disruption of a shell of cold
gas being accelerated by strong radiation pressure, motivated by
models of radiation driven outflows in ultraluminous infrared galaxies. 
We find that when the growth rate of RTI is smaller than acceleration time 
scale, the amount of gas that would be pushed away by the radiation field 
is reduced due to RTI.



\end{abstract}


\section{Introduction}
\label{sec:intro}

The Rayleigh-Taylor Instability (RTI) occurs when a high density fluid
is accelerated by a low density fluid in a gravitational field or net acceleration 
\citep[e.g.][]{Chandra1961, Sharp1984}.
In an astrophysical context, RTI with or without magnetic field has
been studied extensively both numerically
\citep[e.g.][]{Junetal1995,Dimonteetal2004,
  StoneGardiner2007b,StoneGardiner2007a} and experimentally
\citep[e.g.][]{AndrewsSpalding1990,Dalziel1993}.  In most previous studies, support
against gravity was provided by a gradient of the gas pressure 
or magnetic pressure alone. In this paper, we are interested in the dynamics of the RTI
when support against gravity is provided primarily by radiation
pressure. This situation is expected, for example, in the inner region
of accretion disks around compact object; at the interface of an
\ion{H}{2} region produced by massive star clusters and the
surrounding medium \citep[e.g.][]{KrumholzMatzner2009, JacquetKrumholz2011} 
and the dusty tori around luminous Active Galactic Nuclei (AGN) \citep[e.g.][]{Krolik2007,Schartmannetal2010}.  In
ultraluminous infrared galaxies (ULIRGS), radiation pressure on the dust grains
may provide the dominant vertical support against gravity and
accelerate the cold (neutral) gas outflows observed in these systems
\citep[e.g.][]{Murrayetal2005,Thompsonetal2005,Martin2005}. 
The structure of radiation
dominated bubbles around super-Eddington massive stars 
also depends on whether RTI develops on the bubble interface
\citep[e.g.][]{Krumholzetal2009,Kuiperetal2012}. Radiation RTI, together 
with Kelvin-Helmholtz instability, may also be important to generate the 
turbulence in the interstellar medium (ISM) \citep[e.g.][]{MckeeOstriker2007}. 
Thus, understanding the effects of
radiation field on linear growth rate and non-linear structures of RTI, 
as well as assessing the role of radiative RTI in driving turbulence and possibly limiting
the effectiveness of radiative support and driving
\citep{Kuiperetal2011,KrumholzThompson2012},
is an important step to understand those physical systems.

Effects of a radiation field on RTI have been studied analytically in
the linear regime with different approximations. 
In the optically thin limit, radiation flux only provides a background
acceleration, and both \cite{MathewsBlumenthal1977} and
\cite{JacquetKrumholz2011} conclude that stability depends only on the
direction of the effective gravitational acceleration.
\cite{Krolik1977} studied the global RTI for an incompressible slab of
gas accelerated by radiation pressure.  Unstable mode exists if local
radiative acceleration correlates positively with total optical depth.
Recently, \cite{JacquetKrumholz2011} examined the effects of radiation field on
the stability of the interface between two fluids in the isothermal
and adiabatic limits. In the isothermal limit, the radiation flux only
provides an effective background acceleration. In the adiabatic limit,
the radiation pressure plays the same role as gas pressure. However,
the background state they envisage for the linear analysis (two half
infinite constant density planes) cannot precisely be in thermal
equilibrium (see section \ref{bgstate}) with non-zero absorption
opacity.  This may affect the growth rate they calculate and the
evaluation in the non-linear regime when the timescale for thermal
evolution is comparable to or shorter than the growth rate of the RTI.

If there is no density jump in a radiation supported atmosphere
without magnetic field, the nature of the stability criterion is not
completely settled. With the Eddington approximation,
\cite{SpiegelTao1999} and \cite{Shaviv2001} find that there are global
hydrodynamic unstable modes for a hot atmosphere, even for
luminosities moderately below the Eddington limit. However, the
instabilities found by \cite{SpiegelTao1999} are not present in the
\cite{Shaviv2001} analysis.  Also,~\cite{SpiegelTao1999} note that
\cite{Marzec1978} arrived at a different conclusion by solving the
full transfer equation numerically in a PhD thesis.
\cite{Turneretal2005} found overturning modes in simulations with a
fixed temperature at the lower boundary, which are consistent with Shaviv's
Type I modes, although the authors find no evidence for the Type II
modes \citep{Shaviv2001}.

Here we will focus on the traditional case for RTI with a high 
density fluid on top of a low density fluid separated by an infinite thin interface. 
In the previous analytical studies, simplifications are necessary to make the
problem tractable.  For example, radiation pressure is usually assumed
to be isotropic, which is not true in general at the interface. In
this work, we relax these assumptions by solving the radiation
hydrodynamic equations numerically. We study both the linear regime,
using an Eddington tensor computed self-consistently from the
time-independent radiation transfer equation (i.e., we allow for
anisotropic radiation pressure at the interface), and we also follow
the RTI into the non-linear regime which is not possible in analytical
studies.  We note in passing that we have also used our numerical methods to test
the stability of 
a radiation supported atmosphere with a smooth density profile, and find no
evidence for instability, in agreement with \cite{Turneretal2005}.

The paper is organized as follows. In section \ref{sec:equations}, we
describe the equations we solve and the numerical code we use. We then
consider the problem of RTI with a single, initially static interface
between two fluids of different density.  We describe the background
equilibrium state and initial perturbations in section \ref{bgstate},
and summarize our results in section \ref{sec:Result}. In section
\ref{shell}, we describe simulations of RTI in shells being
accelerated by radiation forces.  We summarize and conclude in section
\ref{sec:summary}.


\section{Equations}
\label{sec:equations}
We solve the radiation hydrodynamic equations in the mixed frame with radiation source terms given 
by \cite{Lowrieetal1999}. We assume local thermal equilibrium (LTE) and that the Planck
and energy mean absorption opacities are the same. Detailed discussion of the equations we solve can be found in
\cite{Jiangetal2012}. With a vertical gravitational acceleration $\bg$, the equations are
\begin{eqnarray}
\frac{\partial\rho}{\partial t}+\bfnabla\cdot(\rho \bv)&=&0, \nonumber \\
\frac{\partial( \rho\bv)}{\partial t}+\bfnabla\cdot({\rho \bv\bv+{\sf P}}) &=&\rho\bg-{\bf \tilde{S}_r}(\bP),\  \nonumber \\
\frac{\partial{E}}{\partial t}+\bfnabla\cdot\left[(E+P)\bv\right]&=&\rho\bv\cdot\bg-c\tilde{S}_r(E),  \nonumber \\
\frac{\partial E_r}{\partial t}+\bfnabla\cdot \bF_r&=&c\tilde{S}_r(E), \nonumber \\
\frac{1}{c^2}\frac{\partial \bF_r}{\partial t}+\bfnabla\cdot{\sf P}_r&=&{\bf \tilde{S}_r}(\bP).
\label{dimequation}
\end{eqnarray}
Here, $\rho$ is density, ${\sf P}\equiv P\bI$ (with $\bI$
the unit tensor) and $P$ is gas pressure, $\sigma_a$ and $\sigma_s$ are the absorption and
scattering opacities. Total opacity (attenuation coefficient) is $\sigma_t=\sigma_s+\sigma_a$.  The total gas energy density is
\begin{eqnarray}
E=E_g+\frac{1}{2}\rho v^2,
\end{eqnarray}
where $E_g$ is the internal gas energy density.   We adopt an equation of state
for an ideal gas with adiabatic index $\gamma$, thus
$E_g=P/(\gamma-1)$ and $T=P/R_{\text{ideal}}\rho$, where
$R_{\text{ideal}}$ is the ideal gas constant. 
The radiation pressure ${\sf P}_r$ is related to the radiation energy density $E_r$
by the closure relation
\begin{eqnarray}
{\sf P}_r=\bfr E_r.
\end{eqnarray}
where $\bfr$ is the variable Eddington tensor (VET). Radiation flux is 
$\bF_r$ while $c$ is the speed of light. 
The gravitational acceleration $\bg$ is fixed to be a constant value along the $-z$ axis.  
${\bf\tilde{S}_r}(\bP)$ and $\tilde{S}_r(E)$ are the radiation momentum and energy source terms.

Following \cite{Jiangetal2012}, we use a dimensionless set of equations and variables
in the remainder of this work.  We convert the above set of equations to the dimensionless 
form by choosing fiducial units for velocity, temperature and 
pressure as $a_0, \ T_0$ and $P_0$ respectively. Then 
units for radiation energy density $E_r$ and flux $\bF_r$ are $a_rT_0^4$ 
and $ca_rT_0^4$. In other words, $a_r=1$ in our units. The dimensionless speed of light is $\Crat\equiv c/a_0$. 
The original dimensional equations can then be written to the following dimensionless 
form 
\begin{eqnarray}
\frac{\partial\rho}{\partial t}+\bfnabla\cdot(\rho \bv)&=&0, \nonumber \\
\frac{\partial( \rho\bv)}{\partial t}+\bfnabla\cdot({\rho \bv\bv+{\sf P}}) &=&\rho\bg-\mathbb{P}{\bf S_r}(\bP),\  \nonumber \\
\frac{\partial{E}}{\partial t}+\bfnabla\cdot\left[(E+P)\bv\right]&=&\rho\bv\cdot\bg-\mathbb{PC}S_r(E),  \nonumber \\
\frac{\partial E_r}{\partial t}+\mathbb{C}\bfnabla\cdot \bF_r&=&\mathbb{C}S_r(E), \nonumber \\
\frac{\partial \bF_r}{\partial t}+\mathbb{C}\bfnabla\cdot{\sf P}_r&=&\mathbb{C}{\bf S_r}(\bP),
\label{equations}
\end{eqnarray}
where the dimensionless source terms are,
\begin{eqnarray}
{\bf S_r}(\bP)&=&-\sigma_t\left(\bF_r-\frac{\bv E_r+\bv\cdot{\sf P} _r}{\mathbb{C}}\right)+\sigma_a\frac{\bv}{\mathbb{C}}(T^4-E_r),\nonumber\\
S_r(E)&=&\sigma_a(T^4-E_r)+(\sigma_a-\sigma_s)\frac{\bv}{\mathbb{C}}\cdot\left(\bF _r-\frac{\bv E_r+\bv\cdot{\sf  P} _r}{\mathbb{C}}\right).
\label{sources}
\end{eqnarray}
The dimensionless number $\Prat\equiv a_rT_0^4/P_0$ is 
a measure of the ratio between radiation pressure and gas pressure in the fiducial units. 
We prefer the dimensionless equations because the dimensionless numbers, such as 
$\Crat$ and $\Prat$, can quantitatively indicate the importance of the radiation field  
as discussed in \cite{Jiangetal2012}.

We solve these equations in a 2D $x-z$ plane with the recently developed radiation transfer module in Athena \citep[][]{Jiangetal2012}. 
The continuity equation, gas momentum 
equation and gas energy equation are solved with modified Godunov method, which couples the stiff radiation source terms to the calculations of 
the Riemann fluxes. The radiation subsystem for $E_r$ and $\bF_r$ are solved with a first order implicit Backward Euler method. Details on the 
numerical algorithm and tests of the code are described in \cite{Jiangetal2012}. 
The variable Eddington tensor is computed from angular quadratures of the specific intensity $I_r$, 
which is calculated from the time-independent transfer equation 
\begin{eqnarray}
\frac{\partial I_r}{\partial s}=\kappa_t (S-I_r).
\label{calEdd}
\end{eqnarray}
Details on how we calculate the VET, including tests, are given in \cite{Davisetal2012}. 

\section{Background Equilibrium State}
\label{bgstate}
As is usual, the background equilibrium state used to study RTI in this work is an interface which separates 
two uniform fluids with different densities. 
If there is no radiation field, gravitational acceleration (or effective gravitational acceleration) must be
balanced by a gas pressure gradient. With radiation, the background state needs to 
satisfy both mechanical and thermal equilibrium if the absorption opacity is nonzero. 
Because the background state is uniformly parallel to the interface, we only need to 
consider equation (\ref{equations}) perpendicular to the interface (and parallel 
to the direction of gravity), which is the $z$ direction. 
The equilibrium state should satisfy the following equations
\begin{eqnarray}
\frac{\partial P}{\partial z}+\rho g&=&\mathbb{P}\sigma_t F_r,\nonumber \\
E_r&=&T^4,\nonumber\\
\frac{\partial F_r}{\partial z}&=&0,\nonumber\\
\frac{\partial P_r}{\partial z}&=&-\sigma_t F_r.
\label{balance0}
\end{eqnarray}
The second equation states that radiation temperature must be the 
same as gas temperature in thermal equilibrium when there is a non-zero 
absorption opacity. The third equation means radiation flux 
is a constant.
Combing the four equations, and making the Eddington approximation, 
we find
 \begin{eqnarray}
\frac{\partial}{\partial z}\left(P+\frac{1}{3}\mathbb{P}T^4\right)=-\rho g.
\label{balance1}
\end{eqnarray}
That is the gradient of the sum of gas and radiation 
pressure that balances gravity. 
The gas pressure $P$ is related to the density $\rho$ and temperature $T$ via ideal 
gas equation of state
 \begin{eqnarray}
P\equiv R_{\rm{ideal}}\rho T=\rho T.
\label{RT:static}
\end{eqnarray}
At the interface, if there is a density jump, 
there must be a corresponding jump in the 
gas temperature to satisfy continuity of the total pressure. 
In thermal equilibrium, the radiation temperature 
and gas temperature must be the same, which means 
that radiation temperature will also 
jump at the interface. In this case, the diffusion equation 
cannot be satisfied at the interface. That is, 
with radiation, and if absorption opacity is non-zero, 
this configuration cannot satisfy both mechanical 
and thermal equilibrium because gas 
pressure and radiation pressure depends on 
temperature in different ways. 
Thus an interface with a constant non-zero 
absorption opacity cannot be used as an equilibrium 
background state.

One way of circumventing this constraint is to simply consider a
background that is not strictly in thermal equilibrium, but with a
thermal timescale that is much longer than the growth rate of the
instabilities of interest. One can then assume that the evolution of
the background has only minor effects on the stability properties and
subsequent evolution.  The energy equation in (\ref{equations})
indicates that the timescale to achieve thermal equilibrium can be
reduced by making the flow less relativistic (lower $\Crat$), less
radiation dominated (lower $\Prat$), or by lowering the absorption
opacity.  If we take the limit that absorption opacity goes to zero on
one (or both) sides of the interface, $E_r$ and $T$ decouple and $T$
can be discontinuous while $E_r$ remains continuous.  From this point
on, we consider domains where absorption opacity is zero everywhere,
and the thermal time is effectively infinite.

\subsection{Equilibrium State with Pure Scattering Opacity}
In order to study the evolution of density discontinuities 
due to RTI for radiation pressure supported interface, 
and to compare to previous studies of RTI of interfaces in 
gas pressure supported atmosphere, we 
only consider material with pure scattering opacity. The specific scattering opacity 
$\kappa$ is assumed to be a constant (as in the case of electron scattering opacity). 
Then $\sigma_t=\rho\kappa$, and radiation temperature can be independent 
of gas temperature. 

The background equilibrium state with pure scattering 
opacity is calculated in the following way. For a given density, opacity 
and gravitational acceleration, many equilibrium states can be constructed, 
depends on the relative contributions from the gas pressure and radiation pressure 
to support the gravity. 
We first choose a constant flux $F_{r,0}$ and define
\begin{eqnarray}
\alpha\equiv \frac{\Prat\sigma_t F_{r,0}}{\rho g}.\label{alpha}
\end{eqnarray}
This $\alpha$ is the fraction of gravitation acceleration that is 
balanced by radiation pressure gradient and $1-\alpha$ is the fraction 
balanced by gas pressure gradient. The radiation pressure 
profile can then be calculated from the diffusion equation, and the gas pressure 
profile can be calculated according to $\partial P/\partial z=-(1-\alpha)\rho g$. 
Solutions in the two fluids are matched in the interface by requiring that 
gas and radiation pressure are continuous.

There are two special background equilibrium states. The first case is $\alpha=0$, 
which means that the initial background radiation flux $F_{r,0}$ is zero and gravity is balanced by gas pressure gradient alone. 
This is very similar to previous studies of RTI except that 
now the material is embedded in a uniform radiation field.
The second case is $\alpha=1$, which means that the gas pressure is uniform and gravitational acceleration 
is balanced by radiation pressure gradient alone (Eddington limit). 
We will focus on the two special cases first.

\begin{table}[htdp]
\caption{Parameters of the simulations with only one interface. }
\begin{center}
\begin{tabular}{cccccccc}
\hline
Label & Rad/Hydro & $\kappa$ & $\Prat$ & $\alpha$ & $\rho_+$ &	$\rho_-$	& Perturbation \\
\hline
$A$ & Hydro & ---  & --- & 0 & 4 &  1	& Single \\
$B$ & Rad	 & $1$  & $10^4$ & 1 & 4 & 1 &Single \\
$C$ & Rad	 & $1$  & $1$ & $1$ & 4 &   1 &	Single \\
$D$ & Rad	 & $10^3$  & $10^4$ & $1$ & 4 & 1	& Single \\
$E$ & Rad	 & $1$  & $1$ & $1$ & $16$ & 1	& Single \\
$F$ & Hydro & ---  & --- & 0 & 4 &  1 &	Random \\
$G$ & Rad	 & $1$  & $1$ & $1$ & 4 &  1  &Random \\
$H$ & Rad	 & $1$  & $10^4$ & $1$ & 4 & 	1	& Random \\
$I$ & Rad	 & $10^3$  & $1$ & $1$ & 4 & 	1   &	Random \\
$J$ & Rad	 & $1$  & $1$ & $1$ & $16$ & 	  1  &Random \\
$L$ & Hydro & ---  & --- & 0 & 4 &  1	& eigenvector \\
$M$ & Rad	 & $1$  & $1$ & $1$ & 4 &   1 &	eigenvector \\
$N$ & Rad	 & $1$  & $10^4$ & 1 & 4 & 1 & eigenvector \\
$O$ & Rad	 & $10^3$  & $10^4$ & $1$ & 4 & 1	& eigenvector \\
$P$ & Rad	 & $10^5$  & $10^2$ & $1$ & 4 & 1	& eigenvector \\
\hline
\hline
\end{tabular}
\end{center}
\label{Parameters}
\end{table}%

\subsection{Simulation Setup}
\label{setup}
The 2D simulations are done in the $(x,z)$ plane with gravitational acceleration along 
the $-z$ direction. In our units, the size of the simulation box 
is $(-0.5,0.5)\times(-1,1)$ and a resolution of $128\times512$ grid points is used. 
The gravitational acceleration 
is $g=0.1$, the dimensionless speed of light 
$\Crat$ is $10^4$, while gas pressure $P=10$ and $E_r=10$ at the 
interface $z=0$, except for simulations $L$ --- $P$, where we use $P=1$ and $E_r=3$ 
at the interface. The density in the upper region $z>0$ is denoted by 
$\rho_+$ while it is $\rho_-$ in the bottom half $z<0$. 
The isothermal sound speed in $\rho_-$ is $c_s=1$. 
Time is measured in units of the sound crossing time along the 
horizontal direction. The ratio between the
free-fall velocity to the isothermal sound speed is  $(gL_z)^{0.5}/c_s=0.45$. 

Periodic boundary conditions are used in the horizontal direction, while the 
gradient of both gas and radiation pressure in the vertical direction 
are extended to the ghost 
zones to achieve a better hydrostatic equilibrium. Radiation flux along the $z$ 
direction is fixed to be the initial value to balance the gravity in the ghost zones 
while the $x$ component is continuously copied to the ghost zones. 
The boundary condition of the short characteristic module used to
calculate the VET is set such that a constant flux is also maintained
in that module. Since the divergence of the radiative flux is zero due
to the assumption of negligible absorption opacity, this simply means
we adjust the incoming radiation field at the base of the domain to
provide the appropriate flux at the bottom boundary.  For simplicity
we assume isotropic radiation for the incoming intensity, although the
angular distribution will generally be problem dependent.  This
allows us to ensure that the ratio of radiation flux to radiation energy density
from the radiation transfer module will be consistent with the result obtained
by evolving eqs.~(\ref{equations}). If this consistency is not guaranteed, 
the incorrect VET can cause numerical instability near a stable interface. 

The underlying issue is that the boundary conditions on the short
characteristics solver should be consistent with the initial condition
and boundary condition on the radiation energy and momentum
equations. For example, if the domain is assumed to be embedded in an
optically thick background so that $F_r \ll E_r$ initially, then the
Eddington tensor should be nearly uniform ($\sim 1/3$) everywhere.  If
the boundary conditions on the short characteristics method due not
make the same assumption (e.g. if they assumed the upper boundary
corresponded to vacuum), the VET would increase appreciably from $1/3$
near the upper boundary even though it should be constant.  In this
case, very small variations in $E_r$ can drive large and unphysical
variations in $\bF_r$.  We find that forcing the boundary conditions
in the modules to return a consistent ratio of flux to energy density
ensures a consistent Eddington tensor and ameliorates this problem.


We consider three
sets of perturbations to the initial background.  In the first two sets,
only the vertical velocity is perturbed, either by a single mode in the form 
\begin{eqnarray}
v_z=0.0025(1+\cos(2\pi x))(1+\cos(\pi z)), 
\label{singlemode}
\end{eqnarray}
or by a random collection of modes in the form
\begin{eqnarray}
v_z=v_{z,0}(1+\cos(\pi z)), 
\label{randommodes}
\end{eqnarray}
where $v_{z,0}$ is a random number uniformly 
distributed between $-0.025$ and $0.025$.
We also consider another set of single mode perturbations 
where we perturb $v_x, v_z, P_r, F_{rx}$, and
$F_{rz}$ based on the linear analysis
described in the appendix.  We normalize $v_z$ as
\begin{eqnarray}
(v_{z})_{\pm}=0.01 \cos(2\pi x) \exp(\mp 2\pi z),
\label{singlemode2}
\end{eqnarray}
with perturbations in other quantities chosen to correspond to
the incompressible eigenvectors.
Other parameters of the 
simulations  are listed in Table \ref{Parameters}.  We only
present results for interfaces with $\rho_+ > \rho_-$.  As in the purely
hydrodynamic case, the interface is expected to be stable
when the lower density fluid overlies the higher density fluid.
We have confirmed this numerically for several test cases.

\section{Results}
\label{sec:Result}
In this section, we describe the linear growth and non-linear evolution of RTI 
for different background states. By comparing a set of 
controlled simulations, we explore the behavior
of RTI with different parameters, such as opacity and 
the ratio between radiation pressure and gas pressure 
for the two background states $\alpha=0$ and $\alpha=1$.  
We will first use the Eddington approximation by assuming 
radiation pressure to be isotropic, thereafter $\bfr=1/3\bI$. 
We then relax this assumption and study the effects of anisotropic 
radiation pressure on RTI.



\subsection{Growth rate for a single mode perturbation}
\label{RTinterface}

For a non-radiative, gas pressure supported interface, the linear regime 
has been studied extensively 
\citep[e.g.][]{Chandra1961}.
For a mode with wavelength $\lambda$ and thus wavenumber 
$k\equiv 2\pi/\lambda$, the growth rate $n$ of the inviscid,
incompressible RTI is given by
\begin{eqnarray}
n=\sqrt{gk\frac{\rho_+-\rho_-}{\rho_++\rho_-}}.
\label{growthformula}
\end{eqnarray}
However, for the radiation pressure supported interface, it is not
possible to derive a general, but still simple formula for the growth
rate of the radiation RTI.
Nonetheless, we can still examine the behavior of RTI
in various regimes. For example, \cite{JacquetKrumholz2011} derived
the growth rate of RTI for the radiation supported interface in the
optically thin as well as adiabatic limit. However, it is likely that
their results in the adiabatic limit will be affected by their
non-equilibrium background state. In the optically thin limit, the
background radiation flux plays the role as an effective gravitational
acceleration and so the growth rate will be very similar to the value
given by equation (\ref{growthformula}) with $g$ replaced by the
effective gravitational acceleration $g_{\rm eff}$.  Since $g_{\rm
  eff}=0$ for $\alpha=1$, this would correspond to a marginally stable
state with zero growth rate. In the very optically thick limit when
photons are fully coupled with the gas and radiation field is
isotropic, the radiation pressure behaves in the same way as gas
pressure and so the growth rate will be very similar as given by
equation (\ref{growthformula}). (See the appendix for further
discussion.)  The intermediate regime is the most complicated case, as
the Eddington tensor varies with $z$ and standard linear analysis if
of limited utility.


\begin{figure}[hcp]
\centering
\vspace{-3cm}
\includegraphics[width=0.98\hsize]{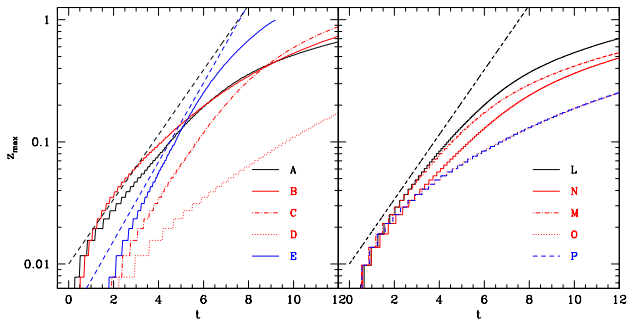}
\vspace{-2cm}
\caption{\emph{Left:} Growth of the vertical displacement $z_{\text{max}}$ between
the top of the low density fluid bubbles and the initial interface due to
RTI for different background states with 
a single mode perturbation given by eq.~(\ref{singlemode}).
Simulation parameters are listed in Table \ref{Parameters}.
The black and blue short dashed lines are the theoretical predictions from
eq.~(\ref{growthformula}) for interfaces with density ratios $\rho_+/\rho_-$
of $4$ and $16$, respectively. \emph{Right:} The same as the left panel, 
except for simulations initially perturbed with the eigenvectors.  
}
\label{RTgrowth_single}
\end{figure}

\begin{figure}[hcp]
\includegraphics[width=0.98\hsize]{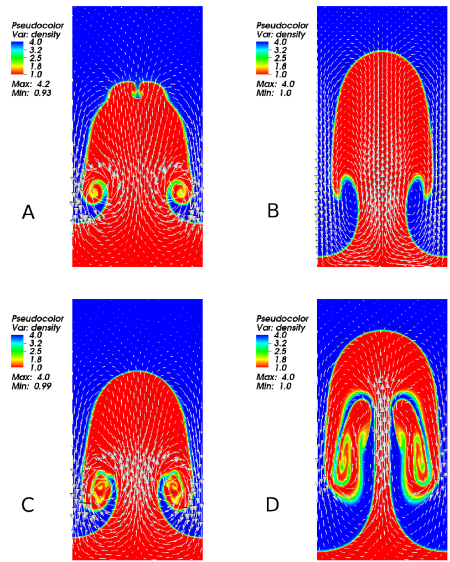}
\vspace{-1cm}
\caption{Density and velocity field for the non-linear regime of RTI 
with a single mode perturbation. From left to right, top to bottom, the 
simulations are $A$ at time $8.8$, $B$ at time $14.0$, $C$ at time 
$11.6$ and $D$ at time $29.3$. Parameters of the simulations 
are listed in Table \ref{Parameters}.  The 
growth history of the four cases are shown in Figure \ref{RTgrowth_single}. 
All the simulations adopt the Eddington approximation $\bfr=1/3\bI$, which 
means the radiation pressure is assumed to be isotropic.
}\label{2DSingleMode}
\end{figure}

In addition to the density ratio at the interface, we anticipate that
the behavior of RTI for radiation supported interface could differ
depending on the background radiation properties, such as the ratio
between radiation pressure and gas pressure and the characteristic
optical depth.  We proceed to vary these parameters and measure the
growth rate from our simulations.  We quantify the growth rate of RTI
using $z_{\text{max}}$, the maximum distance of the interface from
$z=0$.  $z_{\text{max}}$ is measured by the perturbed regions that
have density different from the initial density in those positions by at least $10\%$,
such as the bubbles and fingers.

We first consider the case with single mode eigenvector perturbations.
As discussed in the Appendix, we expect the growth rate consistent with 
eq.~(\ref{growthformula}) since $\Crat \gg 1$ and the Eddington tensor is
diagonal. The results are shown in the right panel of Figure \ref{RTgrowth_single}.  For
a density ratio $\rho_+/\rho_-=4$ and wavenumber $k=2\pi$,
eq.~(\ref{growthformula}) predicts $z_{\text{max}}\propto \exp(0.61 t)$,
which is shown as solid line in the right panel of Figure \ref{RTgrowth_single}.  We find
initial growth is faster than predicted for $z_{\text{max}} \lesssim
0.02$ (corresponding to $\lesssim 5$ cells).  At such early times the
vertical deformation of the interface is not well resolved and
increases in the resolution improve agreement with constant growth
rate solution.  At later times, non-linear effects become important
and the growth rate drops below the prediction for all runs, including
the purely hydrodynamic case.

Although the curves are not uniformly consistent with exponential
growth at a single constant growth rate, the range of growth rates
inferred from the right panel of Figure \ref{RTgrowth_single} 
are close to the prediction
of linear theory.  Furthermore, many aspects of the linear theory are
borne out by the results.  Since $\Prat$ and $\kappa$ do not appear in
eq.~(\ref{growthformula}) the linear theory predicts no dependence on
these parameters to first order in $\Crat^{-1}$.  This is essentially
correct for the initial stages of growth where the solutions remain in
the linear regime.  However, after about one sound crossing time in
the simulations with large $\Prat$ and $\kappa$ ($N$, $O$, and $P$),
the growth rate of $z_{\rm max}$ slows relative to the hydrodynamic
and low $\Prat$ cases.  This appears to be a non-linear behavior as
the amplitude of the perturbation grows.

We attribute this non-linear behavior primarily to the effects of
radiation drag, by which we mean that the fluid motions are damped by
oppositely directed radiation forces.  This requires that velocity
fluctuations (on average) tend to directed opposite to the comoving
radiation flux.  Note that this is not equivalent to radiation
viscosity, which is absent due to our assumption that the radiation
field is isotropic \citep{MihalasMihalas1984}.  To linear order in
perturbed quantities and lowest order in $\Crat^{-1}$, radiation drag
is only relevant for compressible motions.  The effects of radiation
drag are most apparent in simulation where both $\Prat$ and $\kappa$
are large.  Runs $O$ and $P$, with different values $\Prat$ and
$\kappa$ but the same product of $\Prat \kappa$, follow similar
trajectories. This is consistent with the linear analysis since the
characteristic frequency $\nu \propto \Prat \kappa$ controls the
importance of radiation drag on the solution.  See the Appendix for
further discussion.

The results of right panel of Figure \ref{RTgrowth_single} 
can be compared with the
second set of single mode perturbations that are not eigenvectors,
which are also shown in the left panel of Figure \ref{RTgrowth_single}.  In this case, a
similar diversity of growth rates is observed, with notably reduced
growth for the runs with larger $\Prat$ and $\kappa$.  As with the
eigenvector perturbations this trend of slower growth with larger
radiation pressure appears to be a result of the radiation drag.
In this case, the initial perturbations give rise to compressive
motions which are largely absent in the eigenvector initial conditions.
This allows radiation drag effects to become important at lower
amplitude than in the eigenvector runs.  But the range of late time
(non-linear) evolution observed is qualitatively similar in the two
sets of runs.

We have performed another simulation with the same parameters as
simulation $D$, but with $\alpha=0$ i.e. with a background state
supported by a gas pressure gradient instead of a radiation pressure
gradient.  We find a very similar growth history to that in simulation
$D$.  This confirms that the difference in the role of gas and
radiation pressure in supporting the background has no qualitative
effect on the resulting behavior.  Radiation has relatively modest
effects when $\Prat$ is small, but generically leads to slower growth
when $\Prat$ and $\kappa$ are large.  This strengthens the conclusion
that damping of the compressible motions by radiation drag is the main
effect of the radiation field.  The fluid simply has more difficulty
moving with respect to strong radiation field.

\subsection{Non-linear structure with a single mode perturbation}
\label{result:single}
The non-linear structure developed by RTI with single mode 
initial perturbations are shown in 
Figure \ref{2DSingleMode}, taken at late times in simulations $A-D$. 
By comparing the left two panels (simulation $A$ and $C$), we see 
that when the radiation pressure is comparable to the gas pressure and radiation 
pressure is assumed to be isotropic, RTI for the radiation supported 
interface and gas pressure supported interface have very similar non-linear 
structures. This is consistent with the fact that they have very similar 
linear growth phase as shown in Figure \ref{RTgrowth_single}, especially 
when $\rho_+$ is increased to $16$ (simulation $E$).

The non-linear structures for the large radiation pressure cases 
(simulation $B$ and $D$) are quite different from simulations $A$ 
and $C$, especially at small scales. The Kelvin-Helmholtz (KH) 
instability, which is produced during the non-linear 
phase of RTI because of the shearing between the rising bubbles and falling fingers, 
is clearly damped in simulation $B$ where the optical depth across the 
box is $1$. When optical depth is increased to $10^3$ for simulation 
$D$, the small scale structures can survive because the size of those 
structures are much larger than photon diffusion length. However, 
these structures are still quite different when compared to the structures in 
simulations $A$ and $C$. The vortices are strongly stretched along 
the vertically direction when a strong radiation pressure exists.  


\subsection{Non-linear structures of RTI with random initial perturbations}
\label{sec:random}

To study the interactions between different modes of RTI, we add
random perturbations to the vertical velocity as given by equation
(\ref{randommodes}). As in section \ref{result:single}, the effect of
radiation field on the RTI can be shown by comparing a set of
controlled simulations ($F$, $G$, $H$, $I$, $J$), the parameters of
which are listed in Table \ref{Parameters}. The Eddington
approximation is still used in those simulations.


 \begin{figure}[hcp]
\includegraphics[width=0.98\hsize]{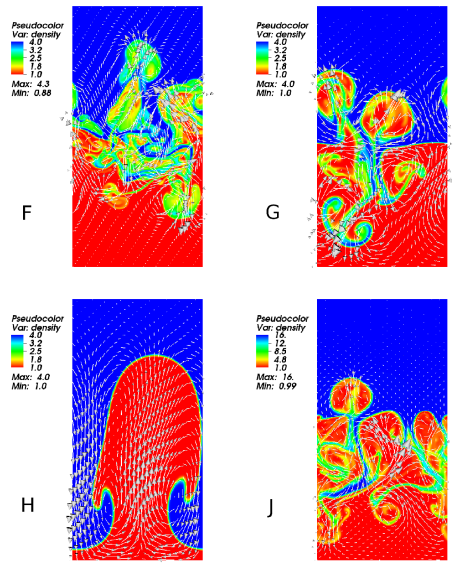}
\vspace{-2cm}
\caption{Density and velocity field for the non-linear regime of RTI
with random initial perturbations.
From left to right, top to bottom, the 
simulations are $F$ at time $15.2$, $G$ at time $15.8$. $H$ at 
time $26.0$. $J$ at time $10.7$. 
All the simulations use the Eddington approximation. Parameters of 
the simulations are listed in Table \ref{Parameters}.}
\label{2DRandom}
\end{figure}

\begin{figure}[hcp]
\centering
\includegraphics[width=0.98\hsize]{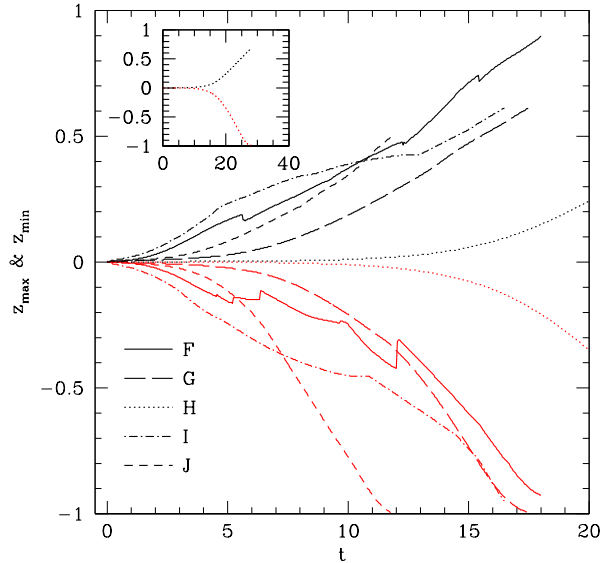}
\caption{Growth of vertical displacements
$z_{\text{max}}$ and $z_{\text{min}}$ (the largest distance between the initial interface and 
the perturbed regions for $z>0$ and $z<0$ respectively) for RTI with random perturbations. 
The black (red) lines are for $z_{\text{max}}$
($z_{\text{min}}$). 
Evolution over a longer time scale 
for simulation $H$ is shown in the left corner. Parameters of the simulations 
are given in Table \ref{Parameters}.
}
\label{RTymax_random}
\end{figure}

In Figure \ref{2DRandom}, we show four snapshots from simulations
$F,G,H,J$ (from left to right and from top to bottom) to illustrate
the range of non-linear structures that we find.  By comparing the top
two panels ($F$ and $G$), we see that when radiation pressure is
comparable to gas pressure ($\Prat=1$), the non-linear outcome of RTI
for the radiation supported interface is similar to the case with gas
pressure supported interface.  There are two major differences when a
radiation field is present. The first is that the dense, heavy fingers
sink faster than the light bubbles rise up.
This is not the case for the non-radiative RTI. Another difference is
that for radiative RTI, the fluid is less mixed. In
non-radiative RTI, there are many cells with density distributed
between the whole range of the maximum and minimum density due to the 
small scale turbulence driven by secondary KH instabiity. However,
with radiation field, there are fewer such cells.  The result when
$\rho_+$ is increased by a factor of $4$ is shown in the bottom right
panel (simulation $J$). The mixing rate is increased as cells with
intermediate densities occupy a larger fraction of the volume. The
non-linear structures in this case bear more similarity to those in
the purely hydrodynamic run. Recall that simulation $E$ has a linear
growth rate very close to the growth rate of normal RTI, as shown in
Figure \ref{RTgrowth_single}.  These results suggest that a higher
ratio of $\rho_+/\rho_-$ reduces the impact of radiation and 
leads to behavior more consistent with the purely hydrodynamic RTI.

The effects of strong radiation pressure ($\Prat=10^4$, simulation H)
on the non-linear structure is shown in the left bottom panel of
Figure \ref{2DRandom}, where optical depth across the box along the
horizontal direction is $1$.  Due to the strong damping effect of the
radiation field, all the small scale perturbations are damped and only
the mode with the longest horizontal wavelength grows. In other words,
the radiation field sets a minimum scale, above which the turbulence
due to RTI can be sustained when radiation pressure is significant.
Then the non-linear outcome is very similar to that in the runs with
single mode perturbation with the same wavelength (simulation $B$
shown in Figure \ref{2DSingleMode}).  The bubbles and fingers are well
defined and mixing ratio is significantly reduced.

As in section \ref{RTinterface}, we use $z_{\text{max}}$ and
$z_{\text{min}}$ to quantify the properties of RTI for the case with
initial random perturbations.  Here $z_{\text{min}}$ is similar to
$z_{\text{max}}$ but only for the bottom half space ($z<0$) while
$z_{\text{max}}$ is only for the top half space ($z>0$).  Growth
history for the five simulations ($F,G,H,I,J$) are shown in Figure
\ref{RTymax_random}.  For non-radiative RTI (simulation $F$), the
structure is symmetric with respect to the initial interface, which
means $z_{\text{max}}$ and $z_{\text{min}}$ grow at approximately the
same rate.  However, this is not true for the RTI when a strong
radiation field is present.  As shown in Figure \ref{RTymax_random},
for the case with radiation supported interface, $z_{\text{min}}$
reaches $-1$ at an earlier time than the time when $z_{\text{max}}$
reaches $1$. The asymmetry is reduced when opacity is
increased (simulation $I$).  The non-monotonic growth of 
$z_{\text{max}}$ and $z_{\text{min}}$ is likely caused 
by the mixing of the cells at the top of the bubbles and fingers. 
When the mixing is reduced (as in simulation H), the 
changes of $z_{\text{max}}$ and $z_{\text{min}}$ are always 
monotonic. 

\begin{figure}[hcp]
\centering
\includegraphics[width=0.98\hsize]{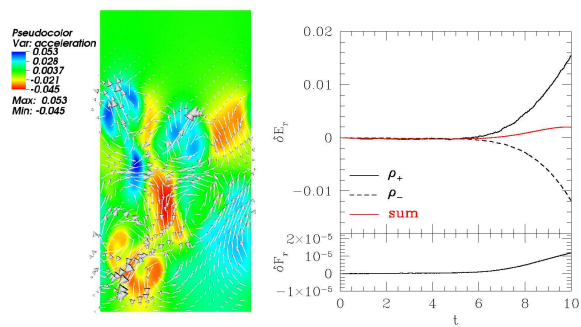}
\vspace{-2cm}
\caption{\emph {Left:} Snapshot of velocity (the vectors) and net acceleration (the color) due to radiation 
and gravity at time $t=15.8$ for simulation $G$. The corresponding density distribution at this time  is shown 
in the top right panel of Figure \ref{2DRandom}. \emph {Right:} The top panel shows 
the temporal evolution of the change of radiation energy density for low density fluid (dashed black line), 
high density fluid (solid black line) and total radiation energy density (red line) 
for simulation $G$. The bottom panel shows the change of the divergence of the flux across the 
whole simulation box compared with the initial value. }
\label{Radacce}
\end{figure}

\begin{figure}[hcp]
\centering
\includegraphics[width=0.7\hsize]{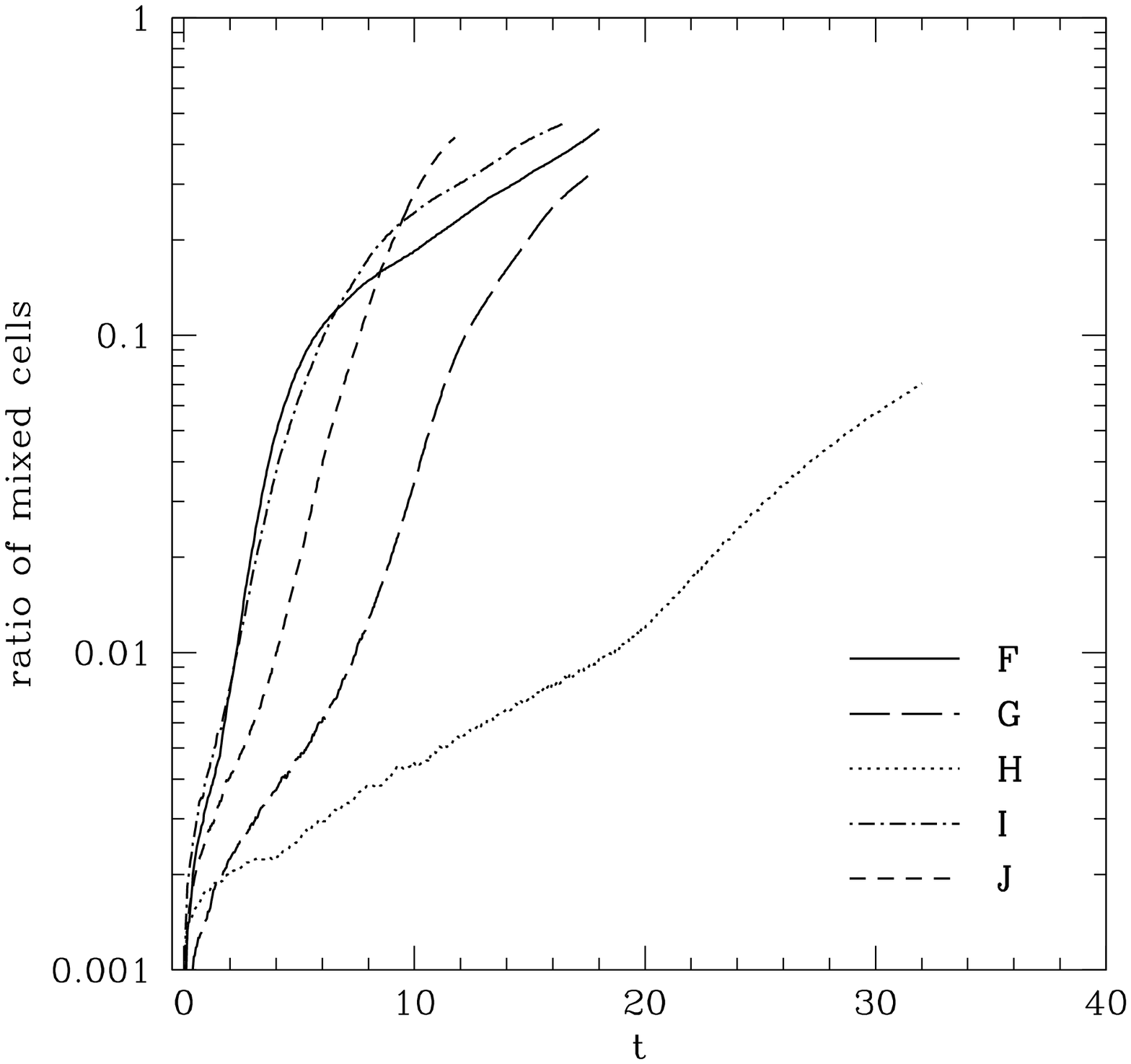}
\caption{Mixing ratio for the simulations with initial 
random perturbations. Parameters of the simulations are 
given in Table \ref{Parameters}.}
\label{RTmix_random}
\end{figure}

In the left panel of Figure \ref{Radacce}, we show the net
acceleration due to gravity and radiation flux along the vertical
direction, $a_{\text{net}}=\Prat\kappa F_{r0,z}-g$, for simulation $G$
at time $15.8$.  The initial hydrostatic equilibrium state has
$a_{\text{net}}=0$ everywhere.  The snapshot shows that for the
bubbles moving up, $a_{\text{net}} > 0$, which means that they are
pushed up by the radiation force. For the falling fingers,
$a_{\text{net}}<0$, which means that radiation force is not able to
balance the gravity.  The right panel of Figure \ref{Radacce} shows
the history of the change of radiation energy density $\delta E_r$ 
and the divergence of the radiation flux across the whole simulation 
box $\delta F_r$ with respect to
the initial values. Depending on whether the density in each cell is larger 
than $0.5(\rho_+ +\rho_-)$ 
or not, we can locate the regions containing high density fluid or low density fluid. 
The radiation energy density inside the high density fluid is
increased because the high density fluid falls from the low radiation
energy region to the high radiation energy region. The situation is
inverted for the low density fluid. Because of the density (and, therefore,
photon mean free path) difference, the radiation energy density of the low density fluid
is changed more quickly than the high density fluid. Thus the bubbles
lose the support from radiation force more quickly than the fingers
and the asymmetry arises. When opacity is increased for the fingers,
the photon mean free path is reduced and the diffusion time is longer.


When density contrast is increased by a factor of $4$ (simulation $J$), the growth 
rate is increased and the asymmetry is decreased. When radiation pressure is 
large enough ($\Prat=10^4$ for simulation H), the growths of $z_{\text{max}}$ and 
$z_{\text{min}}$ are significantly decreased, which is consistent with what we find 
for the single mode perturbation shown in section \ref{RTinterface}.

\begin{figure}[hcp]
\includegraphics[width=0.7\hsize]{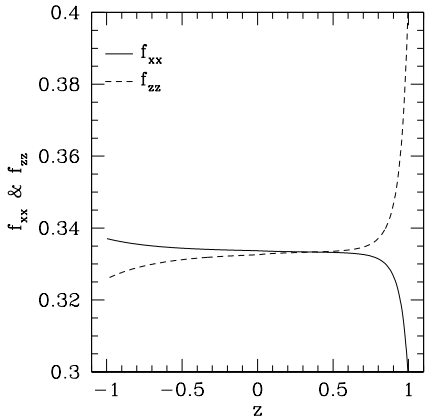}
\caption{Vertical profile of the initial $x-x$ ($f_{xx}$) and $z-z$ ($f_{zz}$) 
components of the Eddington tensor 
for the simulation G3 described in section \ref{sec:Eddeffect}. 
The Eddington tensor is invariant along the $x$ axis due 
to the symmetry of the initial condition. $f_{xx}$ decreases 
and $f_{zz}$ increases monotonically from the bottom to the top.}
\label{Eddnormal}
\end{figure}

\subsection{Mixing of the two fluids in the Rayleigh-Taylor Instability}
\label{sec:mixing}
With random perturbations, the non-linear regime of the RTI generates 
turbulence, in which there is significant mixing.

To quantify the degree of mixing, we count the number of cells with
density between $90\%$ of the maximum density $d_+$ and $1.1$ times
the minimum density $d_-$.  The degree of mixing can be estimated
by the ratio between the mixed cells and the total number of
cells. The mixing ratio for the simulations with initial random
perturbations is shown in Figure \ref{RTmix_random}.  For simulation
$G$ with $\kappa=1$ and comparable gas pressure and radiation
pressure, the mixing ratio is only $71\%$ of the mixing ratio for
non-radiative RTI.  When we increase the opacity to $\kappa=10^3$
(simulation $I$), the mixing ratio is very close to the case of
non-radiative RTI because the photons are so tightly coupled to the
fluid.  When all the available modes are optically thick,
the mixing that results from RTI is very similar in the radiative and 
non-radiative cases. However, if  the smallest eddies are
optically thin while the largest modes are optically thick, photons will
decouple from the fluid on small scales and there will
be less mixing on these scales.  When radiation pressure is
increased to $\Prat=10^4$, the mixing ratio is dropped to $\sim
8\%$. This is consistent with the fact that all the small scale
structures are damped and KH instability,
which is the most important cause of mixing the
two fluids in normal RTI, is suppressed by the strong radiation field.
We caution that we have not performed a numerical convergence study of mixing
in the radiation RTI, and the precise mixing fractions we quote in this section are
likely to change with numerical resolution.  In fact, without explicit viscosity,
converged results for the mixing rate are not possible.
Nonetheless, the trends in the mixing rate
with radiation pressure and opacity reported in this section should be independent of the
viscosity and therefore numerical resolution.

\subsection{Effects of Anisotropic Radiation Pressure}
\label{sec:Eddeffect}

All the simulations listed in Table \ref{Parameters} adopt the
Eddington approximation for the radiation field.  Since the radiation
pressure is assumed to be isotropic, these results only apply to very
optically thick flows.  We now consider cases with moderate to low
optical depth and take into account the anisotropy of the radiation
field by calculating VET self-consistently with equation
(\ref{calEdd}).  This allows us to assess the impact of anisotropic
radiation pressure on the development of the RTI. Here we focus on the
cases with a constant background radiation flux.

We first carry out a simulation G2 with Eddington approximation, 
which has almost the same setup as 
simulation G with a different initial perturbation. 
Here we perturb the density randomly as 
\begin{eqnarray}
\rho=\rho_0+\delta\rho(1+\cos(\pi z)), 
\label{randommodesrho}
\end{eqnarray}
where $\delta\rho$ is a random number uniformly distributed 
between $-0.5\rho_0$ and $0.5\rho_0$ and the perturbation is only 
applied when $|z|<0.5$. The initial radiation energy density at the bottom 
is $1.72$ and decreases to $0.22$ at the top due to the 
background radiation flux $F_{r,0}$, which balances the gravity. 
All the other initial conditions are 
the same as simulation G. A third simulation employing the
VET is labeled as G3. Incoming specific intensities are applied at both 
the top and bottom of the box to maintain the same 
constant background flux $F_{r,0}$. 

The initial vertical profile of the Eddington tensor 
is shown in Figure \ref{Eddnormal}. 
For the pure scattering opacity case, the anisotropy 
is mainly caused by the stratification of the radiation 
energy density in order to provide the background 
radiation flux. From bottom to top, $f_{xx}$ 
decreases while $f_{zz}$ increases monotonically. 
The Eddington tensor changes vary rapidly within roughly 
one optical depth from the boundary and is close to $1/3$ 
in the middle of the simulation box. If we increase the vertical 
size of the simulation box, the profile of the Eddington 
tensor within one optical depth from the boundary remains 
the same while the middle region will become more isotropic.  
Not that we do not find $f_{xx}=f_{zz}=1/3$ at the bottom boundary
because the downward going (backscattered) radiation field is not
perfectly isotropic.

\begin{figure}[hcp]
\vspace{-3cm}
\includegraphics[width=1.0\hsize]{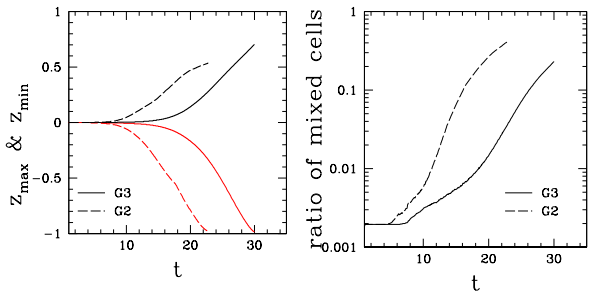}
\vspace{-3cm}
\caption{Effects of VET on the growth rate and mixing ratio 
for RTI in a radiation supported interface. 
The left panel shows the evolution of $z_{\text{max}}$ 
and $z_{\text{min}}$ as in Figure \ref{RTymax_random} 
while the right panel shows the mixing ratio as in 
Figure \ref{RTmix_random}.
Simulation G2 adopts the Eddington approximation,
while a VET is calculated self-consistently for 
simulation G3. All the other initial conditions 
are the same for the two simulations. }
\label{Eddeffect_growth}
\end{figure}

\begin{figure}[hcp]
\includegraphics[width=0.9\hsize]{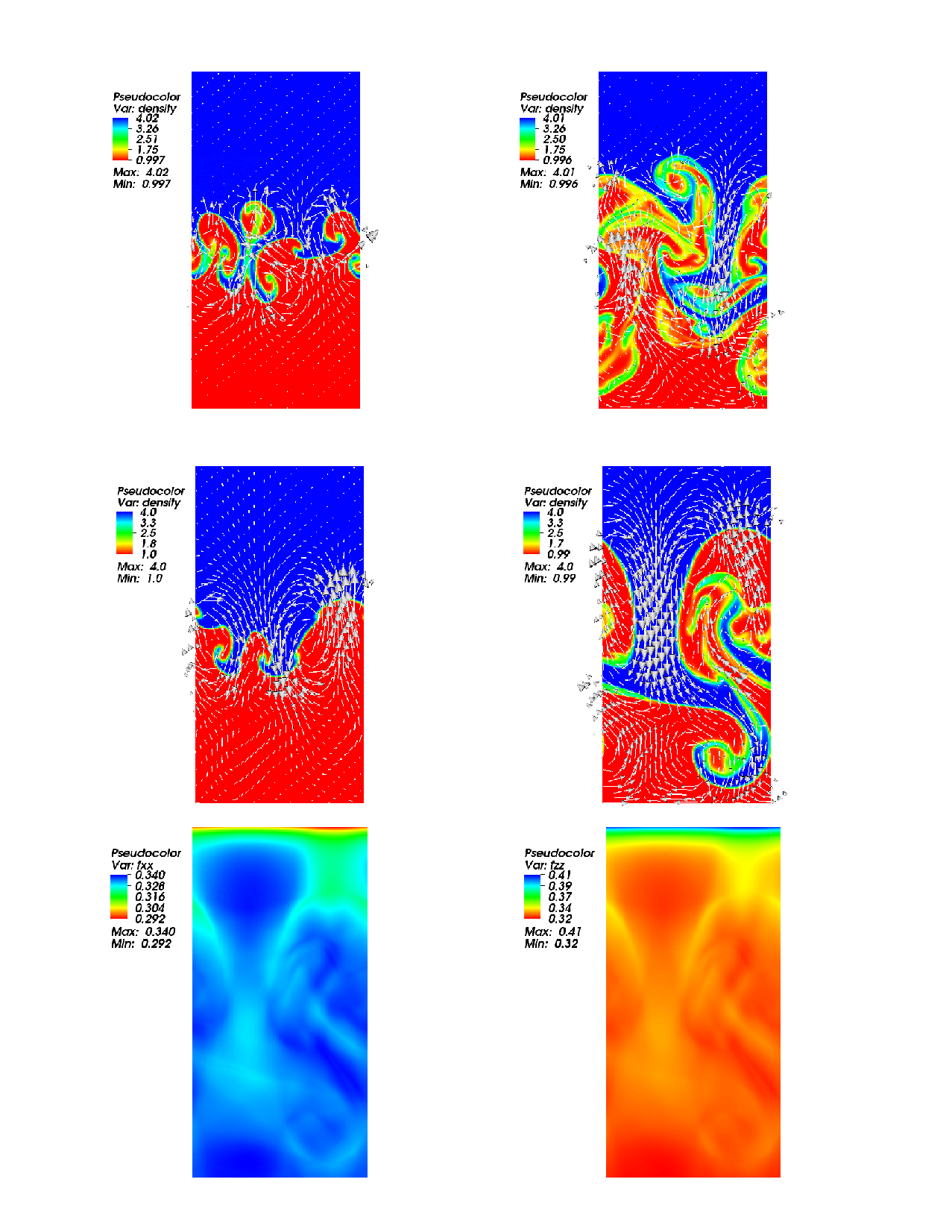}
\caption{
Comparison of the non-linear structure 
from simulations G2 and G3, showing the 
effect of anisotropy in the radiation field captured through
use of a VET.  \emph {Top:}
Density distribution for G2 at time $15.2$ (left) 
and $21.6$ (right). \emph {Middle:} 
Density distribution for G3 at time $21.6$ (left) 
and $28.9$ (right).  The 
vectors in the those panels are the radiation 
flux with the background flux subtracted. 
\emph {Bottom:}
The horizontal ($f_{xx}$) and vertical ($f_{zz}$) components of 
the VET for simulation G3 at time $28.9$.  }
\label{VETComp}
\end{figure}

The growth history and mixing ratio of the two simulation G2 and G3
are shown in Figure \ref{Eddeffect_growth}. It is interesting to see
that growth rate of G3 with VET is slower than the growth rate of
G2. The mixing ratio from G3 is also smaller than the mixing ratio
from G2.  In Figure \ref{VETComp}, we show snapshots of the non-linear
structures to illustrate the differences between the two simulations.
Consistent with Figure \ref{Eddeffect_growth}, there are more mixed
(intermediate density) cells from simulation G2 and fewer in
simulation G3.  The top and middle panels in Figure \ref{VETComp} also show
that the characteristic scale of eddies in G3 is larger than that of
simulation G2.  The slower growth rate in G3 is consistent with
equation (\ref{growthformula}), which predicts that the growth rate of
RTI decreases with decreasing wave number.

Components of the Eddington tensor $f_{xx}$ and $f_{zz}$ at the same
time for simulation G3 are also shown in the two bottom panels of
Figure \ref{VETComp}. Interestingly, $f_{xx}$ and $f_{zz}$ actually
have similar structures to the density distribution, which is also the
photon mean free path distribution. There is a clear transition for the Eddington
tensor between the fingers and bubbles: $f_{xx}$ is enhanced inside
the finger while $f_{zz}$ is decreased compared with the initial
values. Considering the complicated behavior of the Eddington tensor with position,
it is not surprising that the non-linear structures produced by the RTI are different
for simulations with VET and the Eddington approximation.
Moreover, this result is a cause for concern regarding the reliability of results
computed with much more approximate methods, such as flux-limited diffusion.

\section{Radiation Supported Cold Shell}
\label{shell}
In ULIRGS, radiation from OB association is usually super-Eddington
for adjacent dusty gas
\citep[e.g.][]{Murrayetal2005,Thompsonetal2005,KrumholzMatzner2009}.
The radiation field from these stars may have
important feedback effects on the interstellar medium, such as
reducing the star formation efficiency
\citep[e.g.][]{Thompsonetal2005} and driving the cold
neutral outflows seen in many of these systems \citep{Martin2005}.
Momentum driving may be more effective than thermal driving at
accelerating clouds to escape velocities while not overheating and
ablating the cold neutral gas
\citep[e.g.][]{Murrayetal2005,Martin2005,Socratesetal2008}.

It has been recently argued \citep{KrumholzThompson2012} that the RTI
limits the ability of radiation pressure to support ULIRGs against
gravity due to the temperature sensitivity of the dust opacity.  Here
 we consider a related but somewhat different question of whether
radiation pressure can effectively accelerate cold shells of gas before
the RTI grows to a point that it limits radiative driving or provides
enough mixing that the shell is no longer cold relative to the ambient
medium.

Here we give a simplified two dimensional model to demonstrate the
effects of RTI on the gas in the interstellar medium. Initially, a
dense cold shell is in gas pressure equilibrium with surrounding hot
gas. Gravity is balanced by the radiation pressure gradient.  We
assume pure constant scattering opacity so that we can start from an
equilibrium state.  Note that real dust can both absorb and scatter
photons, with absorption opacity comparable to or slightly larger than
scattering opacity at the temperature relevant for ULIRGS.

We adopt the parameters characteristic for ULIRGS
\citep[e.g.][]{Thompsonetal2005,Murrayetal2005}, subject to some
constraints on the dynamical range that can be efficiently evolved by
our simulation methods.  The number density of the cold medium is
$n_c=1000$ cm$^{-3}$, which corresponds to a mass density $\rho_c \sim
3.3\times 10^{-21}$ g cm$^{-3}$. The density ratio between the cold
and hot medium is $100$. Temperature of the cold medium is $T_c=50$
K. The cold and hot gas are in gas pressure equilibrium.
Gravitational acceleration is assumed to be $g=10^{-8}$ cm/s$^2$.
A typical size scale is chosen to be $l_0\sim 1$ pc. 
With these parameters, the sound crossing time for $l_0$
is $\sim 10^6$ yr. To see the effects of varying the
opacity, we try two different scattering opacities $1$ cm$^{2}$
g$^{-1}$ and $100$ cm$^{-2}$ g$^{-1}$ so that the optical depths
across $l_0$ for the cold gas, which is labeled as $\tau_s$, are
$0.01$ and $1$ respectively.  Background radiation flux is chosen such
that acceleration due to radiation force is $1.01$ times the
gravitational acceleration. Thus, the cold shell will be accelerated
upwards, representing the dusty gas which is pushed away by the strong
radiation field. Taking $\rho_c, T_c$ and $l_0$ as density,
temperature and length units in our simulation, the dimensionless
parameters $\Prat=8.1\times 10^3$ and $\Crat=7.10\times 10^5$. 

The size of the simulation box is $l_0\times 4l_0$ and the cold shell
is located at the middle, which extends from $-0.5l_0$ to
$0.5l_0$. Periodic boundary conditions are used along the horizontal
direction.  For the vertical directions, density is fixed to be
$0.01\rho_c$ in the ghost zones. The vertical components of the
velocity and radiation flux in the ghost zones are fixed to the
initial values while the horizontal components are copied from the
last active zones. This vertical boundary condition can maintain the
constant radiation flux entering and leaving the simulation box.
Numerical resolution is $128\times 512$ for all the simulations.

\begin{figure}[hcp]
\includegraphics[width=0.8\hsize]{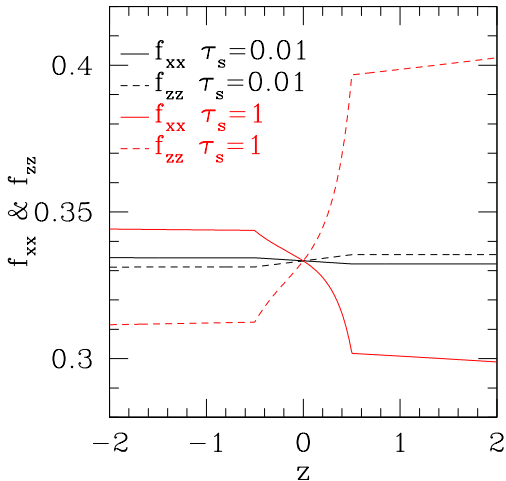}
\caption{
Initial profiles of the $x-x$ and $z-z$ components of the Eddington
tensor for the radiation supported cold shell shown in Section
\ref{shell}. The red lines are for the case when the optical depth
across the cold shell is $1$ while the black lines are for the case
with opacity $100$ times smaller. The rapid change in the Eddington
tensor at positions $z=-0.5$ and $z=0.5$ are due to the jump in
density (and, therefore, photon mean free path) at the interfaces.  Although the density
contrasts are $100$ for both the red and black lines, the change of
Eddington tensor is much larger for $\tau_s=1$.}
\label{ShellEdd}
\end{figure}

\begin{figure}[hcp]
\includegraphics[width=0.98\hsize]{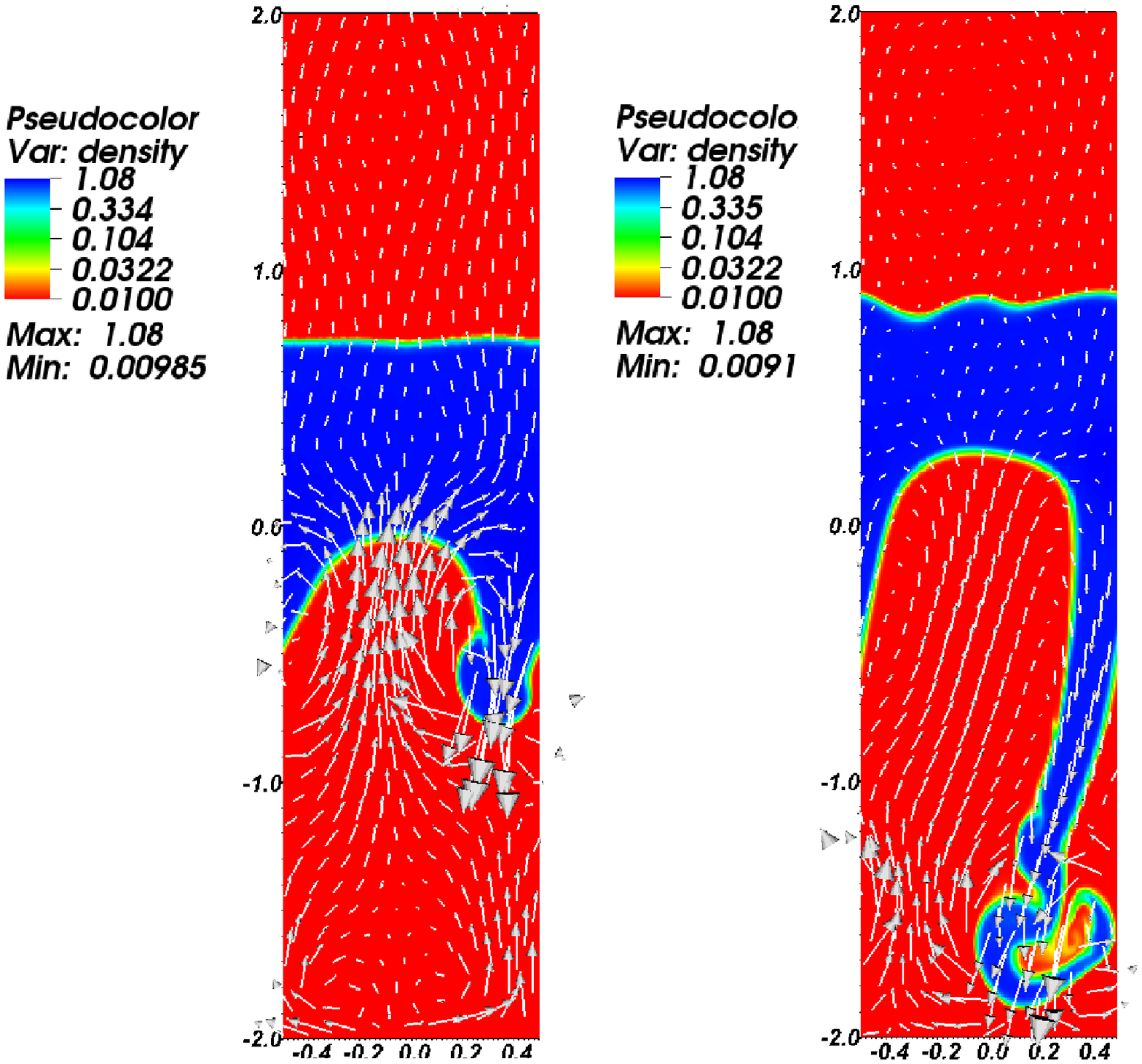}
\caption{Structures of the cold shell at times $2.3\times 10^6$ yr (left panel) and $1.1\times 10^7$ yr (right panel). 
Initial optical depth across the cold shell is $\tau_s=1$. Fingers and bubbles due to RTI are formed, 
which allows the cold dense gas to fall downwards instead of being pushed away by the background radiation 
flux. The vectors are the velocity field.}
\label{Shellthick}
\end{figure}

As discussed in section \ref{sec:Eddeffect}, to calculate the VET
correctly, we find it is necessary to make the ratios of radiation
flux to radiation energy density from the radiation transfer modules
to be consistent with that obtained by evolving
eqs.~(\ref{equations}).  The boundary conditions on the radiative
transfer model enforce a constant flux and assume that the incoming
intensity at the lower boundary is isotropic.

Given the density distribution and opacity, the initial VET is shown
in Figure \ref{ShellEdd} for the two cases $\tau_s=0.01$ and
$\tau_s=1$. When $\tau_s=1$, the cold shell is optically thick while the
hot ambient medium is optically thin. Thus we see a rapid change of
Eddington tensor near the interfaces at $z=\pm0.5l_0$. The change is
much smaller when $\tau_s=0.01$ and the radiation field remains nearly
isotropic, consistent with the isotropy of the incoming radiation at the
base of the domain. The vertical component of the Eddington
tensor $f_{zz}$ increases with height while the horizontal component
$f_{xx}$ decreases with height. Anisotropy of the radiation field is larger
when $\tau_s=1$ and much smaller when $\tau_s=0.01$.

\begin{figure}[hcp]
\includegraphics[width=0.98\hsize]{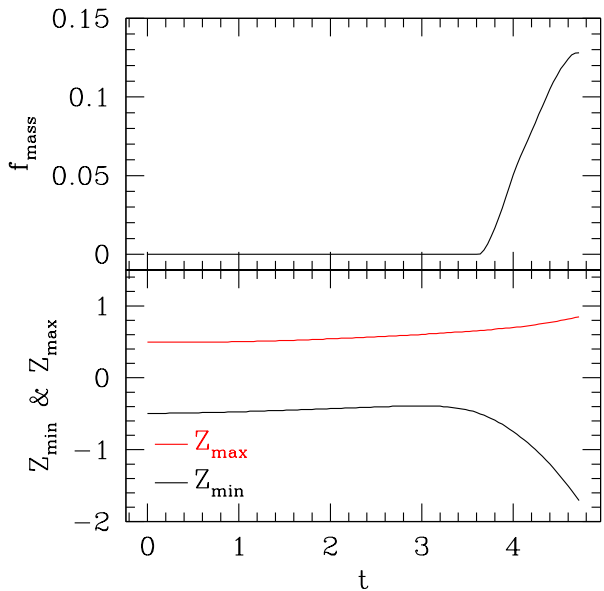}
\caption{\emph{Top:} History of the fraction of cold dense gas that is below the line $-0.5l_0$. Time unit is 
$2.3\times10^6$ yr. Before the RTI develops, no cold dense gas is below the line $-0.5l_0$ because the 
cold gas is pushed away by the radiation field. After RTI grows significantly, over $13\%$ of cold gas falls 
back during $\sim 2.8\times10^6$ yr. \emph{Bottom:} Change of the maximum and minimum heights
of the cold dense gas with time. During the time $t<3.6$, the whole dense shell moves upwards. Once 
RTI grows significantly, some of the cold gas almost falls back to the bottom of the simulation box.    
}
\label{Shellhistory}
\end{figure}

Initially, we perturb the density randomly according to equation
\ref{randommodesrho}.  The random number $\delta\rho$ is uniformly
distributed between $-0.005$ and $0.005$.  As the acceleration due to
the photons is $1\%$ larger than the gravitational acceleration in
this setup, the whole shell will move upwards while RTI
develops. The structure of the cold shell at times $2.3\times 10^6$ yr
and $1.1\times10^7$ yr for the case $\tau_s=1$ are shown in Figure
\ref{Shellthick}. RTI develops quickly at the lower interface, where
the high density medium is on top of the low density medium.  The
finger makes the cold dense gas fall back towards the low density gas
exponentially, instead of being pushed away by the photons. In the top
panel of Figure \ref{Shellhistory}, we show how the fraction of cold
gas that is below the line $-0.5l_0$, $f_{\text{mass}}$, change with
time. Initially all the cold gas is located between $-0.5l_0$ and
$0.5l_0$ and the whole shell is being pushed upwards before RTI
develops, therefore $f_{\text{mass}}$ remains zero when
$t<3.6$. After RTI grows significantly, $f_{\text{mass}}$ increases to
$13\%$ after just $2.8\times10^6$ yr. The maximum and minimum heights of
the cold dense gas, $z_{\text{max}}$ and $z_{\text{min}}$, are shown
at the bottom panel of Figure \ref{Shellhistory}. Consistently with
the history of $f_{\text{mass}}$, the whole shell moves upwards
together during the time $t<3.2$. After $t=3.2$, $z_{\text{min}}$
decreases dramatically instead of increasing with time. The cold gas
eventually reaches the bottom of the simulation box.

\begin{figure}[hcp]
\includegraphics[width=0.85\hsize]{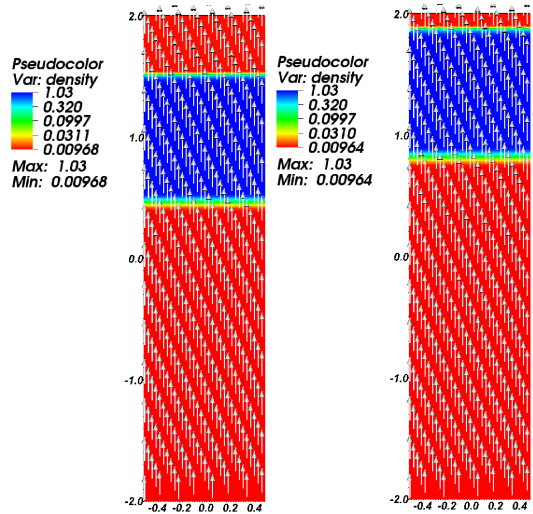}
\caption{The same as Figure \ref{Shellthick} except that the initial optical depth across the cold 
shell is $\tau_s=0.01$ in this case. The two interfaces are both stable in this case and the whole 
cold shell is pushed upwards.}
\label{Shellthin}
\end{figure}

To see the effects of opacity, we decrease the opacity by a factor of $100$ 
so that $\tau_s=0.01$. In Figure \ref{Shellthin}, we show the density and velocity 
distribution at times $2.3\times 10^6$ yr and $1.1\times 10^7$ yr for this case. The evolution of 
the cold shell is quite different in this case compared with the case $\tau_s=1$, in 
that the two interfaces are both stable during the time when the shell crosses the 
simulation box. The whole shell is just pushed away 
by the photons and no cold gas is left.

By comparing Figure \ref{Shellthick} and Figure \ref{Shellthin}, we
can also see the different effect of radiation force in the optically
thick and thin regimes. Initially in both cases, a constant
radiation acceleration, which is larger than the gravitational
acceleration, is applied across the shell to push it upwards. In the
optically thin regime, the acceleration due to the radiation force is
almost unchanged and it behaves like an effective gravitational
acceleration.  However, in the optically thick regime when RTI develops,
the radiation flux is enhanced inside the bubbles and reduced inside
the fingers.  In this case the whole shell no longer moves with a constant
acceleration. Instead, the cold gas falls back quickly through the
fingers.

It is notable that the characteristic timescale for accelerating the
shell $\sim \sqrt{2 l_0/a}$, with $a \sim 0.01 g$ is about $10^7$
years for the parameters under consideration.  Since, the timescale
for appreciable growth of the RTI is a few $\times 10^6$ years when
$\tau_s \sim 1$, there is sufficient time for the RTI to grow and
disrupt the shell. We have also performed runs with the radiation force
is double the gravitational force, but keeping $\tau_s \sim 1$.  This
increases the effective acceleration by a factor of $\sim 100$ and
reduces the timescale for acceleration to $\sim 10^6$ years.  In this
case the shell is accelerated efficiently and reaches the upper
boundary of the domain before the RTI has time to grow appreciable.
Hence we find that efficient acceleration of gas requires low
optical depths or very super-Eddington radiation fluxes.

\section{Summary and Discussions}
\label{sec:summary}

Using our recently developed radiation hydrodynamic module for Athena,
we have performed a set of numerical simulations to see the effects of
strong radiation field on the development of the RTI for a background
state with pure scattering opacity. In many respects, the radiative
RTI is rather similar to the purely hydrodynamic RTI.  Instability is
always present when a high density fluid overlies a lower density
fluid and we find growth rates that are within an order of magnitude
of the hydrodynamic case, even when radiation pressure exceeds gas
pressure by several orders of magnitude.  Nevertheless, we find
that the presence of radiation modified the development of the RTI
in several significant ways.

For a gas pressure supported interface, the presence of a
radiation field will generally reduce the growth rate of RTI, and
growth rate decreases as the radiation pressure increases.  This is
because radiation acts as drag force when the material tries to move
with respect to the radiation field. When radiation pressure is large ($\Prat \sim \Crat$),
the development of small scale structure is also suppressed by the
radiation field.

We find similar effects for a radiation pressure supported interface.
In optically thick limit with an isotropic radiation field, the growth
rate of the radiative RTI is similar to the non-radiative case.  As we
increase the radiation pressure and the opacity we find lower growth
rates, which we again attribute to radiation drag effects.  If the
optical depth across the domain is low, so that the radiation becomes
anisotropic, we generally find the growth rate is reduced relative to
the isotropic case for the same parameters.

For a radiation supported interface with random perturbations, we find
that it is easier for the dense fingers to sink than the light bubbles
to rise up in contrast to the hydrodynamic RTI, where rising and
sinking is nearly symmetric.  This effect is more significant when
opacity is decreased and the diffusion time across the fingers is
reduced.  The mixing process is also suppressed in radiation RTI,
especially when radiation pressure is much larger than gas pressure.

The degree of radiation damping depends both on the opacity and the
ratio between radiation pressure and gas pressure. When radiation
pressure is large ($\Prat\sim\Crat$), all the modes will be affected
by the radiation damping effect significantly as long as photons and
gas are not decoupled for this mode. Turbulence due to RTI associated
with secondary instabilities (e.g. Kelvin-Helmholtz) is significantly
suppressed. When radiation pressure is just comparable to the gas
pressure, only the small scale structures will be damped.

We have used a background state with pure scattering opacity 
because a background state with non-zero
absorption opacity and an interface with a density discontinuity
cannot generally obey both mechanical and thermodynamic equilibrium.
This configuration will be particularly problematic in optically thick
regime and the background state will evolve quickly because of the
thermalization process. Any results obtained for systems with
absorption opacity \citep[e.g.][]{JacquetKrumholz2011} will only be
valid when the growth time of RTI is sufficiently shorter than the
thermalization time or any other short timescale associated with the
interface evolution (e.g. the evolution of an ionization front
in an HII region).

In order to assess the effectiveness of radiative driving of cold gas
outflows we simulate the acceleration of a cold, dense shell, for
which radiation forces exceed gravity. We find that when the optical
depth in the cold shell is $\sim 1$ and the radiation field is just
slightly above Eddington, the RTI can develop quickly and prevent the
acceleration of an appreciable amount of cold gas.  However, if the
radiation field is substantially super-Eddington, or if cold gas is
very optically thin so that the growth rate of RTI is significantly
reduced, the shell can be accelerated efficiently before the onset of
the RTI.  These simulations are too simplified to draw definite
conclusions on the impact of RTI on radiation feedback in star forming
environments, but suggest that the RTI is likely to be important in
some cases. Those simulations also imply that if the 
growth time scale of RTI is smaller than the typical acceleration time 
scale, RTI is likely to change the structure of the system significantly. 
As we find that the growth rate of RTI is similar to normal RTI 
when radiation pressure is not significant ($\Prat\ll \Crat$) and 
the growth rate will be reduced when radiation pressure is very 
significant ($\Prat\sim \Crat$), our results give a general criterion 
on when RTI will be important when a specific astrophysical system 
is considered. 
 Future work will benefit from the use of more realistic
dust opacity and the use of non-grey (i.e. multiple frequency bins)
radiative transfer to differentiate the impact of UV, mid-infrared,
and far-infrared photons. 

One caveat of our simulations is that they are done in 2D. It is known
that RTI leads to different saturation states in 3D compared with 2D
\citep[e.g.][]{Marinaketal1995}.  We plan to explore these effects on
specific problems in future work. However, the linear growth rate is
expected to be unchanged in 3D. We expect that the qualitative
effects we find for the radiation field on the growth of the RTI, such
as the importance of radiation drag in very radiation dominated
regimes, will also be unchanged in 3D.

Other avenues for future research include studying the effects of
absorption opacity on radiation RTI. If the thermal time scale is much
longer than the growth time of RTI, we anticipate that absorption
opacity will have minor effects on the growth and saturation of
RTI. When the thermal time scale is much shorter than or comparable to
the growth time of RTI, the effective equation of state of the fluid
will be modified.  Furthermore, the background state will also change
dramatically in this case, which can also affect the saturation of
RTI.  The effects of absorption opacity on the RTI in specific
astrophysic environments (e.g. with dust opacity in ULIRGS) is a topic
of ongoing research that will be presented in future work.  It is also
important to include magnetic fields and examine the effects of
radiation on magnetic RTI. Our radiation MHD code can also be used to
study RTI in more specific astrophysical systems, such as \ion{H}{2}
regions with massive star formation, photosphere of accretion disks
and supernova remnants.

\section*{Acknowledgments}
We thank Julian Krolik, Jeremy Goodman, Omer Blaes, Norm Murray and 
Emmanuel Jacquet for helpful discussions on the radiation Rayleigh-Taylor 
instability. We also thank the 
referee for valuable comments to improve the paper. This work
was supported by the NASA ATP program through grant NNX11AF49G, and by
computational resources provided by the Princeton Institute for
Computational Science and Engineering.  This work was also 
supported in part by the U.S. National Science Foundation, grant 
NSF-OCI-108849. 
Some computations were
performed on the GPC supercomputer at the SciNet HPC
Consortium. SciNet is funded by: the Canada Foundation for Innovation
under the auspices of Compute Canada; the Government of Ontario;
Ontario Research Fund - Research Excellence; and the University of
Toronto. SWD is grateful for financial support from the Beatrice D. Tremaine 
Fellowship.

\appendix
\section{Linear Analysis}

We present a simplified linear analysis of the Rayleigh-Taylor
Instability in the case of a single interface between two constant
density media with radiation, but in the limit of negligible
absorption opacity.  The background for this problem is described in
Section \ref{bgstate} and specified by eqs.~(\ref{balance0}) -
(\ref{alpha}).  We will consider the case with purely radiation
support ($\alpha=1$) so the both density and gas pressure are constant
within each half plane. The background is assumed to be static with
${\bF_r}=F_0 {\bf \hat{z}}$ and absorption opacity $\sigma_s=\kappa
\rho$.  With these assumptions, hydrostatic equilibrium reduces to
$\mathbb{P} F_0 \kappa = g$.  We consider the 2D problem with a
background that is uniform in x. Variables above and below the
interface will be denoted by subscripts $+$ and $-$, respectively. We
focus on the case with a diagonal Eddington tensor:
$f_{xx}=f_{zz}=1/3$ and $f_{xz}=0$.  Note that this assumption precludes
radiation viscosity although still allows effects due to radiation
drag \citep{MihalasMihalas1984}.

We begin by linearizing eqs.~(\ref{equations}) with perturbations of
the form $f(x,z,t) = A(z) \exp\left[i(k x -\omega t)\right]$.  We
denote the $x$ and $z$ components of the velocity by $u$ and $w$,
respectively. We denote the (scalar) radiation pressure by $P_r$ and
the components of $x$ and $z$ components of $\bF$ by $F_x$ and $F_z$,
respectively.  Constant background quantities are denoted by a
subscript zero, e.g.  background density and vertical flux are
denoted by $F_0$ and $\rho_0$.  We then have:
\begin{eqnarray}
- i \omega \delta\rho + \rho\left(i ku + \odif{w}{z}\right)  & = & 0, \label{eq:mpert}\\
-i\omega \rho u + i k a^2 \delta \rho  + -\Prat \sigma_s \delta F_x + \Prat \frac{4 \sigma_s P_r }{\Crat} u & = & 0,\label{eq:pxpert}\\
-i\omega \rho w + a^2 \odif{\delta \rho}{z} -\Prat \sigma_s \delta F_z + 
\Prat \frac{4 \sigma_s P_r }{\Crat} w & = & 0,\label{eq:pzpert}\\
\frac{-3 i\omega}{\Crat} \delta P_r +i k \delta F_x + \odif{\delta F_z}{z}+\frac{\sigma_s F_0}{\Crat} w & = & 0,\label{eq:erpert}\\
\frac{-i\omega}{\Crat} \delta F_x + i k \delta P_r + \sigma_s \delta F_x - \frac{4 \sigma_s P_r }{\Crat}u  &=& 0,\label{eq:fxpert}\\
\frac{-i\omega}{\Crat} \delta F_z + \odif{\delta P_r}{z} + \sigma_s \delta F_z + \kappa F_0 \delta \rho -
\frac{4 \sigma_s P_r }{\Crat} w &=& 0.\label{eq:fzpert}
\end{eqnarray}
Since the radiation field does not exchange any {\it
  thermal} energy with the gas when $\sigma_a = 0$, gas energy
equation simply reduces to a statement that gas pressure perturbation
are adiabatic $\delta P=a^2 \delta \rho$, with $a^2=\gamma P_0/\rho_0$.
Note that since both $P_0$ and $\rho_0$ are constant, $a$ is a constant.

We further simplify the problem by selectively dropping terms of order
$\Crat^{-1}$.  In particular, we drop the time derivative terms in
eqs.~(\ref{eq:erpert})-(\ref{eq:fzpert}) and the fourth term on the
rhs of eq.~(\ref{eq:erpert}).  These terms are always small relative
to other terms when $\Crat$ is large (i.e. the flow is
non-relativistic).  We nevertheless retain the last terms on the rhs
of eqs.~(\ref{eq:fxpert}) and (\ref{eq:fzpert}) because they
include $P_r$, which can be much larger than $F_0$ in the limit of
large optical depth. It is useful to first define a characteristic frequency
\begin{displaymath}
\nu = \frac{4 P_r g}{F_0 \Crat} = \frac{ 4 P_r \kappa \Prat}{\Crat},
\end{displaymath}
noting that this frequency generally varies with height, due to the
linear dependence on the background $P_r$ on $z$.

We can use equations~(\ref{eq:pxpert}) and (\ref{eq:fxpert}) to solve
for $u$ and $\delta F_x$, obtaining
\begin{eqnarray}
u = \frac{k}{\omega} \left(\frac{g \delta P_r}{\sigma_s F_0} + a^2 \frac{\delta \rho}{\rho} \right),
\end{eqnarray}
and
\begin{eqnarray}
\frac{\delta F_x}{F_0} = -i \frac{k(\omega+i \nu)}{\omega} \frac{\delta P_r}{\sigma_s F_0} + \frac{\nu a^2 k^2}{g \omega^2} \frac{\delta \rho}{\rho}.
\end{eqnarray}

Inserting these into the above equations yields (after some algebra) a system of equations
\begin{displaymath}
\odif{X}{z}=A X
\end{displaymath}
with
\begin{displaymath}
X^{\rm T}=\left(\frac{\delta P_r}{ F_0},\frac{\delta F_z}{F_0},
\frac{\delta \rho}{\rho_0}, \xi\right)
\end{displaymath}
and
\begin{eqnarray}
A=
\left(\begin{array}{cccc}
0 & -\sigma_s & -\sigma_s & -i\frac{\sigma_s \omega \nu}{g}\\
-\frac{k^2(\omega + i \nu)}{\sigma_s \omega} & 0 & -i \frac{k^2 \nu a^2}{\omega g} & 0 \\
0  & \frac{g}{a^2} & 0 & \frac{\omega(\omega + i \nu)}{a^2}\\
\frac{k^2g}{\sigma_s \omega^2} & 0 & \frac{k^2 a^2}{\omega^2} - 1 & 0
\end{array}\right).\label{eq:matrix}
\end{eqnarray}
Here we have replaced $w$ by the Lagrangian displacement using the
definition $w=-i \omega \xi$.  These equations represent for ODEs, but
with non-constant matrix coefficients i.e. the terms with $\nu$ vary
with $z$.  In what follows we will ignore this dependence and assume
constant coefficients.  Since $P_r = P_0 - g \rho_0 z$, $\nu$ is
approximately constants for lengthscales $l \ll \Prat P_0/(g \rho_0)$,
Hence, it is a reasonable approximation for
the subset of simulations with $\Prat \gg 1$.

With the above assumption, solutions take the form
$\mathbf{\chi}^i_{\pm} \exp{(\lambda^i_{\pm} z)}$ with four
eigenvalues $\lambda^i_{\pm}$. The eigenvalues for this system are
given by
\begin{displaymath}
\left(\lambda^2 - k^2 \right)\left(\lambda^2 - k^2 + \frac{\omega(\omega+i\nu)}{a^2}\right)=0.
\end{displaymath}
The first set of eigenvalues $\lambda^2=k^2$ have eigenvectors with $\delta P= a^2 
\delta \rho =0$ and 
\begin{eqnarray}
\frac{\delta F_z}{F_0} & = & -\frac{\omega(\omega + i \nu)}{g} \xi,\nonumber\\
\Prat \delta P_r & = & \frac{\omega^2 \rho}{\lambda} \xi.\nonumber
\end{eqnarray}
These correspond to incompressible modes.  The other set of
compressible eigenvalues have $\lambda \pm \sqrt{k^2-\omega(\omega+i
  \nu)/a^2}$ and has eigenvectors with
\begin{eqnarray}
\delta P & =& \frac{\rho \omega^3(\omega + i \nu)}{g k^2 +\omega^2 \lambda},\label{eq:dpc}\nonumber\\
\Prat \delta P_r & = & \frac{\rho (g \omega^2 \lambda - i \omega^3 \nu)}{g k^2 +\omega^2 \lambda}\label{eq:dprc},\nonumber\\
\delta F_z & = &-F_0\frac{k^2\omega(\omega + i \nu)}{g k^2 + \omega^2 \lambda}\label{eq:dfzc}.\nonumber
\end{eqnarray}

We now apply standard boundary conditions, the first being that the
perturbations vanish far from the interface.  The second set of
conditions is given by matching solutions across the interface.  In
addition to the continuity of $w$, the non-trivial matching conditions
come from integrating the vertical conservation relations in
eqs.~(\ref{equations}).  As \cite{JacquetKrumholz2011} discuss, these
are equivalent to the continuity of the Lagrangian perturbations
across the interface, which for our background correspond to
\begin{eqnarray}
\Prat \Delta P_r & = &\Prat \delta P_r - \rho g \xi,\nonumber\\
\Delta P & = & \delta P,\nonumber\\
\Delta F_z & = & \delta F_z\nonumber.
\end{eqnarray}
Defining perturbations above and below the interface proportional to
\begin{eqnarray}
\chi^{\rm I}_\pm\exp{(\mp k z)} + \chi^{\rm C}_\pm \exp{(\mp \lambda_\pm z)},\nonumber
\end{eqnarray}
where $\lambda_{\pm}$ correspond to the compressible eigenvalues and
are both defined to have positive real parts. These matching
conditions define a matrix $A \mathbf{X}=0$ with $\mathbf{X}^{\rm
  T}=\left(\chi^{I}_+,\chi^{C}_+,\chi^{I}_-,\chi^{C}_-\right)$, and
\begin{eqnarray}
A=
\left(\begin{array}{cccc}
1 & 1 & -1 & -1\\
0 & \frac{\rho_+}{\theta_+} & 0 & \frac{\rho_-}{\theta_-}\\
1 & \frac{k^2 g}{\theta_+} & -1 & - \frac{k^2 g}{\theta_-}\\
\left(\frac{n^2}{gk}-1\right)\rho_+ &  -\frac{\rho_+}{\theta_+} 
\left(g k^2 + \frac{i \omega^3 \nu}{g}\right) & 
\left( \frac{n^2}{gk}+1\right)\rho_- & 
\frac{\rho_-}{\theta_-} \left(g k^2+ \frac{i \omega^3 \nu}{g}\right)
\end{array}\right)\nonumber
,
\end{eqnarray}
where we have defined $\theta_{\pm} = g k^2 \pm n^2 \lambda_\pm$ and $n\equiv - i \omega$.  The
dispersion relation is obtained by setting $\det{A}= 0$, which yields
\begin{eqnarray}
\left(\lambda_+ \rho_- + \lambda \rho_+ \right)
\left(-\frac{\omega^2}{g k} (\rho_+ +\rho_-) - (\rho_+ - \rho_-) \right)=0.\nonumber
\end{eqnarray}
The quantity in the first set of parentheses has a real part which is 
always greater than zero.  Hence if we look for purely imaginary solutions
with growth rates $n\equiv - i \omega$, we obtain
\begin{eqnarray}
n^2 = g k \frac{\rho_+ - \rho_-}{\rho_+ + \rho_-}\nonumber
\end{eqnarray}
which is equivalent to eq.~(\ref{growthformula}), the standard result
for the inviscid, incompressible hydrodynamic problem.  

The robustness of this form for the dispersion relation may seem surprising,
but is straightforward to understand.  First note that the solution
corresponds to the case where $\chi^{C}_\pm =0$ so the
solution on either side of the interface is a purely proportional to
the incompressible eigenvector.  This result stems from the fact that
only the purely incompressible eigenvectors permit the continuity of
both $\delta F_z$ and $\delta P$ across the interface.  In the limit
that $\delta \rho = 0$ and dropping the same $\Crat^{-1}$ terms
as above, eqs.~(\ref{eq:mpert})-(\ref{eq:fzpert}) reduce to
\begin{eqnarray}
iku + \odif{w}{z} = 0,\nonumber\\
-i\omega \rho u + i k \Prat \delta P_r = 0,\nonumber\\
-i\omega \rho w +  \Prat \odif{\delta P_r}{z} = 0,\nonumber
\end{eqnarray}
which resemble the hydrodynamics case, but with gas pressure perturbations
replaced by radiation pressure.  This should be contrasted with the
optically thin case discussed by \cite{JacquetKrumholz2011}, where
there is no $\delta P_r$ since the radiation field is decoupled
from local conditions near the interface.  In that case, since gravity
is exactly balanced by radiation forces the effective gravity is zero
and the interface is nominally stable.

Finally, we note that this result seems to imply that radiation drag have no
effect on the linear growth rate.  However, when $\nu$ is large ($\nu
\gg \sqrt{g k}$), radiation drag forces make the linear regime of
growth be rather limited.  For large values of $\nu$, non-linear
effects can become important, even for very modest velocity
amplitudes.  Indeed our numerical simulations indicate that non-linear
effects become important at lower amplitudes for parameters
corresponding to large $\nu$, as discussed in section \ref{RTinterface}.

\bibliographystyle{apj}
\bibliography{RadRT}
 
\end{document}